\documentclass[twocolumn,prl,amsmath,amssymb,showpacs,superscriptaddress,floatfix]{revtex4-1}
\usepackage{graphicx}
\usepackage{bm}
\usepackage{graphicx}

\renewcommand{\phi}{\varphi}

\newcommand{\ppsi}{{\bf\Psi}}
\newcommand{\pphi}{{\bf\Phi}}

\begin{document}

\title{Trapped two-dimensional condensates with synthetic spin-orbit coupling}
\author{Subhasis Sinha}
\affiliation{\mbox {Indian Institute of Science Education and Research-Kolkata, Mohanpur, Nadia 741252, India.}}
\author{Rejish Nath}
\affiliation{\mbox{Max Planck Institute for the Physics of Complex Systems, N{\"o}thnitzer Strasse 38, D-01187 Dresden, Germany}}
\author{Luis Santos}
\affiliation{\mbox{Institut f\"ur Theoretische Physik , Leibniz Universit\"at, Hannover, Appelstrasse 2, D-30167, Hannover, Germany}}

\date{\today}

\begin{abstract}
We study trapped 2D atomic Bose-Einstein condensates with spin-independent interactions 
in the presence of an isotropic spin-orbit coupling, showing that a rich  
physics results from the non-trivial interplay between spin-orbit coupling, confinement and inter-atomic 
interactions. For low interactions two types of half-vortex solutions with different winding occur,  
whereas strong-enough repulsive interactions result in a stripe-phase similar 
to that predicted for homogeneous condensates. Intermediate interaction regimes are characterized for large 
enough spin-orbit coupling by an hexagonally-symmetric phase with a triangular lattice of density minima 
similar to that observed in rapidly rotating condensates.
\end{abstract}

% \pacs{67.85.-d, 34.50.Cx}

\maketitle

% Introduction
% 

% Spin-orbit coupling

\paragraph{Introduction.} 
The engineering of synthetic electromagnetism in ultra-cold gases
has recently attracted a major attention~\cite{Dalibard2010}. Although an homogeneous 
effective magnetic field may be generated by simple rotation, recent techniques based 
on appropriate laser arrangements and tailored dressed states 
allow for more flexible control of effective artificial 
gauge fields~\cite{Lin2009,Lin2009b}. Interestingly, these techniques allow as well for the  
creation of non-Abelian gauge fields~\cite{Ruseckas2005,Osterloh2005}, and more specifically 
spin-orbit coupling (SOC)~\cite{Juzeliunas2008, Stanescu2008, Liu2007, Sau2011, Campbell2011}, a 
crucial effect in solid-state physics, essential for topological insulators~\cite{Hasan2010}.
Recent ground-breaking experiments 
have explored this fascinating possibility, demonstrating the creation of SOC 
with equal Rashba and Dresselhaus strengths~\cite{Lin2011}.

%%%%%%%%%%%%%%%%%%
%% FIGURE 1
\begin{figure}[ht]
%\vspace{-0.7cm}
\begin{center}
\includegraphics[width=0.5\textwidth,angle=0]{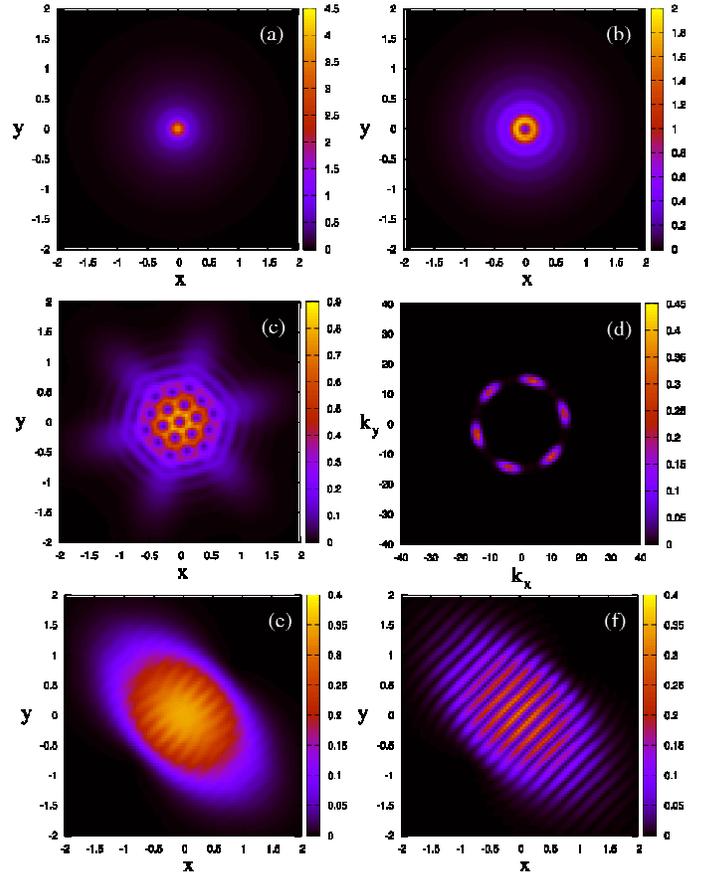}
%\vspace{-0.6cm}
\caption{Total density for $\kappa=15$ and $g=0.05$~(a), $g=0.1$~(b), $g=0.85$~(c), and 
$g=2.0$~(e). Figure~(d) depicts the momentum distribution for $g=0.85$, and 
Fig.~(f) shows the spatial distribution of component $1$ for $g=2.0$.}
\label{fig:1}
\end{center}
\vspace*{-0.6cm}
\end{figure}

%%%%%%%%%%%%%%%%%%%%%%%%%%%%%%%%%%%%%%%%

The creation of SOC in spinor gases opens fascinating questions about the physics of 
ultra-cold gases with SOC, which have aroused a rapidly-growing theoretical attention 
both in what concerns degenerated fermions~\cite{Sau2011,Stanescu2007,Yu2011} 
and Bose-Einstein condensates(BECs)~\cite{Stanescu2008,Wu2008,Merkl2010,Larson2010,Wang2010,Yip2011,
Zhang2011,Xu2011,Kawakami2011}. In particular, Wang et al. have recently shown that the ground-state of 
an homogeneous, i.e. untrapped, two-component BEC with SOC is a single plane-wave phase or 
a spin stripe phase depending on spin-dependent interactions~\cite{Wang2010}. On the other hand, 
non-interacting trapped spin-orbit coupled BECs are expected to present a half-quantum vortex 
configuration~\cite{Stanescu2008,Wu2008}. 

In this Letter, we show that the interplay between trap energy, SOC and interactions leads to 
a rich ground-state condensate physics. We consider in particular the complete 
phase diagram of trapped two-dimensional condensates with spin-independent interactions in the presence of isotropic spin-orbit coupling. This phase diagram is characterized by four different phases. As expected, for low interactions, a half-vortex solution (HV(1/2) below), with angular momentum $l$ such that $|l+1/2|=1/2$, is recovered. 
Increasing interaction leads to a second half-vortex solution (HV(3/2)) 
with a higher $|l+1/2|=3/2$. For a sufficiently large SOC the HV(3/2) phase enters for larger interactions into an hexagonally-symmetric 
phase characterized by the appearance of a triangular lattice of minima in the total density, similar to a vortex lattice 
in fastly rotating BECs~\cite{Raman2001}. Finally, for even larger interactions, 
a spin-stripe phase develops, similar to that found in homogeneous BECs~\cite{Wang2010}. 
We discuss the existence and main properties of these phases.

% Model: description of the system, GPE equation with gauge

\paragraph{Model.} 
We consider a BEC of atoms with mass $m$, confined in two-dimensions on the $xy$ plane, by a tight harmonic confinement along $z$. The atoms have two available internal states, which in typical experiments 
are, as mentioned above, atom-light dressed states of Zeeman sublevels with a proper laser configuration.
Typically, inter-atomic interactions between Zeeman components are very similar.  
Hence, for simplicity we consider spin-independent interactions characterized by the 2D coupling constant $\tilde g>0$, which hence determines as well the interactions between the atoms in the dressed states.  

Both components are equally confined in an isotropic harmonic potential $V(r)=m\omega^2 r^2/2$, with $r^2=x^2+y^2$. We consider that an appropriate laser configuration is chosen such that the atoms experience an isotropic gauge field ${\bf A}=\hbar\tilde\kappa\left (\sigma_x{\bf e}_x+\sigma_y{\bf e}_y\right )$, where $\sigma_{x,y}$ are the Pauli matrices and 
${\bf e}_{x,y}$ are the unit vectors along the directions $x$ and $y$~\cite{Dalibard2010}. 
The physics of the condensate in the presence of SOC and interactions is described by the Hamiltonian
\begin{equation}
\hat H \!\! =\! \! \int \! d^2 r \ppsi^\dag \! \left [\frac{1}{2m}\! \left ( -i\hbar\nabla\! -\!{\bf A} \right )^2 \! + \! V(r)\! + \! \frac{\tilde g}{2}\! \ppsi^\dag\! \cdot\! \ppsi
\right ]\! \ppsi,
\label{eq:GPE}
\end{equation}
where $\ppsi$ is a two component spinor.

% 
% Homogeneous space (review): the stripe
% 

\paragraph{Homogeous solution.}
In the absence of trapping, $V(r)=0$, the condensate minimizes the energy by spontaneously breaking the rotational symmetry, developing a spatial spin modulation along an arbitrary direction~\cite{Wang2010}. Chosing that direction as $x$, we may re-write 
$\ppsi=e^{i(\tilde\kappa x)\sigma_x}\pphi$. For a constant $\pphi$ (with $|\pphi|^2=n$), both the interaction energy, $\tilde g n$, and  
$\langle \left (-i\hbar\nabla - {\bf A} \right )^2 \rangle=0$, acquire its minimum value. Chosing $\pphi^T=\sqrt{n}(1,0)$, 
the spin modulation acquires the form:
\begin{equation}
 \ppsi=\sqrt{n}
\left( 
\begin{array}{c}
\cos(\tilde\kappa x) \\ 
i\sin{\tilde\kappa x} 
\end{array} 
\right)
\end{equation}

% Effects of the trap: effective momentum representation-> mexican hat, effective ``angular momentum''
% 

The presence of the trap modifies and enriches the ground-state physics since it 
introduces an additional energy scale, $\hbar\omega$. 
The trapped BEC is hence best 
described by two dimensionless parameters, $\kappa\equiv \tilde\kappa l_{HO}$, with $l_{HO}^2=\hbar/m\omega$, 
and $g\equiv \tilde g m/\hbar^2$. Note that due to the 2D nature of the problem, 
the dimensionless coupling constant $g$ is actually independent of $\omega$. 
Below we employ dimensionless expressions, using oscillator units.

\paragraph{HV(1/2) Phase.} 
The non-interacting case, $g=0$, is best described 
in momentum representation~\cite{Stanescu2008}. In the absence of trap, the Hamiltonian 
acquires the form $(\hbar{\bf k}-{\bf A})^2/2$, which presents two eigen-energy branches, 
$\epsilon_{\pm}(k)=(k\pm\tilde\kappa)^2/2$, with $k=|{\bf k}|$. The corresponding eigen-spinors are of the form
\begin{equation}
{\bf u}_\pm(\varphi)=\frac{1}{\sqrt{2}}\left( 
\begin{array}{c} 
1 \\ 
\mp e^{i\varphi} 
\end{array} 
\right),
\end{equation}
with $\varphi$ the polar angle of the vector ${\bf k}$. Assuming a dominant SOC, such that $\kappa^2/2$ 
is larger than any other energy scale in the problem, we may 
reduce the analysis to the lowest energy branch: $\ppsi({\bf k})=\psi({\bf k}){\bf u}_-(\varphi)$.
Introducing the trapping potential, $V=-\nabla_k^2/2$, with $\nabla_k$ the gradient in momentum space, 
and using $\psi({\bf k})=\sum_l k^{-1/2} f_l(k) e^{il\varphi}$, one obtains:
\begin{equation}
-\frac{1}{2}\left ( \frac{d^2}{dk^2}f_l-\frac{(l+1/2)^2}{k^2}f_l\right )+\frac{(k-\tilde\kappa)^2}{2}f_l=E_l(k) f_l
\end{equation}
Note that the original kinetic energy term becomes a mexican-hat-like 
``potential'' in momentum space, whereas the trap results in a 
"radial kinetic energy" term and a "centrifugal barrier"~\cite{Stanescu2008}. For $\kappa\gg 1$, 
we may develop around $k\simeq\kappa$, obtaining the eigenenergies
\begin{equation}
E_{nl}=\frac{(l+1/2)^2}{2\kappa^2}+n+1/2,
\end{equation}
where $n$ characterizes the radial excitations of the mexican-hat. 

Without interactions the lowest energy is given by $n=0$ and $|l+1/2|=1/2$, which are states 
of the form:  
\begin{equation}
\ppsi_{l=0,-1}(k)\propto e^{-(k-\tilde \kappa)^2 l_{HO}^2/2} e^{il\varphi}{\bf u}_-, \\
\end{equation}
These states present a spatial dependence of the form:
$\ppsi_0^T({\bf r}=(r,\alpha))\sim (J_0(\kappa r),e^{i\alpha}J_1(\kappa r))$, and 
$\ppsi_{-1}^T({\bf r})\sim (-e^{-i\alpha}J_1(\kappa r),J_0(\kappa r))$, and hence 
constitute half-vortex solutions~\cite{Wu2008}. In the following we denote this phase  
as HV(1/2). Solutions with $l=0$ and $l=-1$ are degenerated,
being eventually selected by quantum fluctuations beyond our mean-field approach. 
However, any linear combination of $l=0$ and $l=-1$ 
produces the same rotationally symmetric total density profile~(Fig.~\ref{fig:1}(a)). The selection of $l=0$ or $l=-1$ would lead to a 
ring-like momentum distribution with $k\simeq\kappa$, whereas linear combinations result in 
a $\cos{\varphi}$ modulation along the ring. 

\paragraph{HV(3/2) Phase.} 
For $\kappa^2\gg 1$, angular excitations along the ring 
are much less energetic than radial excitations. Hence weak interactions result 
in the population of higher $|l+1/2|$ values.
Simulations of Eq.~(\ref{eq:GPE}) show that 
the HV(1/2) phase remains the ground state for sufficiently small $g$~(Fig.~\ref{fig:1}(a)). 
However, the total density of the HV(1/2) solution has a maximum at the trap center due to the 
$J_0(\kappa r)$ contribution. On the contrary, 
the solutions with $n=0$, $|l+1/2|=3/2$, depend as  $J_{l=-2,1}(\kappa r)$ and $J_{l=-1,2}(\kappa r)$,  
and hence have a minimum of the total density at the trap center, presenting 
a reduced interaction energy compared to HV(1/2).  
Solutions with $|l+1/2|=3/2$ become energetically favorable at $g=g_{cr}$, 
when the interaction energy balances the energy difference between both solutions for $g=0$:
\begin{equation}
g_{cr}=\frac{2\pi}{\kappa^2}\left [ \frac{f_2(1,\kappa)}{f_1(1,\kappa)}-\frac{f_2(0,\kappa)}{f_1(0,\kappa)} \right ] 
\label{eq:gcr}
\end{equation}
with $f_\alpha=\int 2rdr e^{-\alpha r^2} \left [J_m^2(\kappa r)+J_{m+1}^2(\kappa r) \right ]^\alpha$. 
For large $\kappa$, $g_{cr}\simeq 2.35 (2\pi/\kappa^2)$. 

\begin{figure}%[ht]
%\vspace{-0.7cm}
\begin{center}
\includegraphics[width=0.35\textwidth,angle=0]{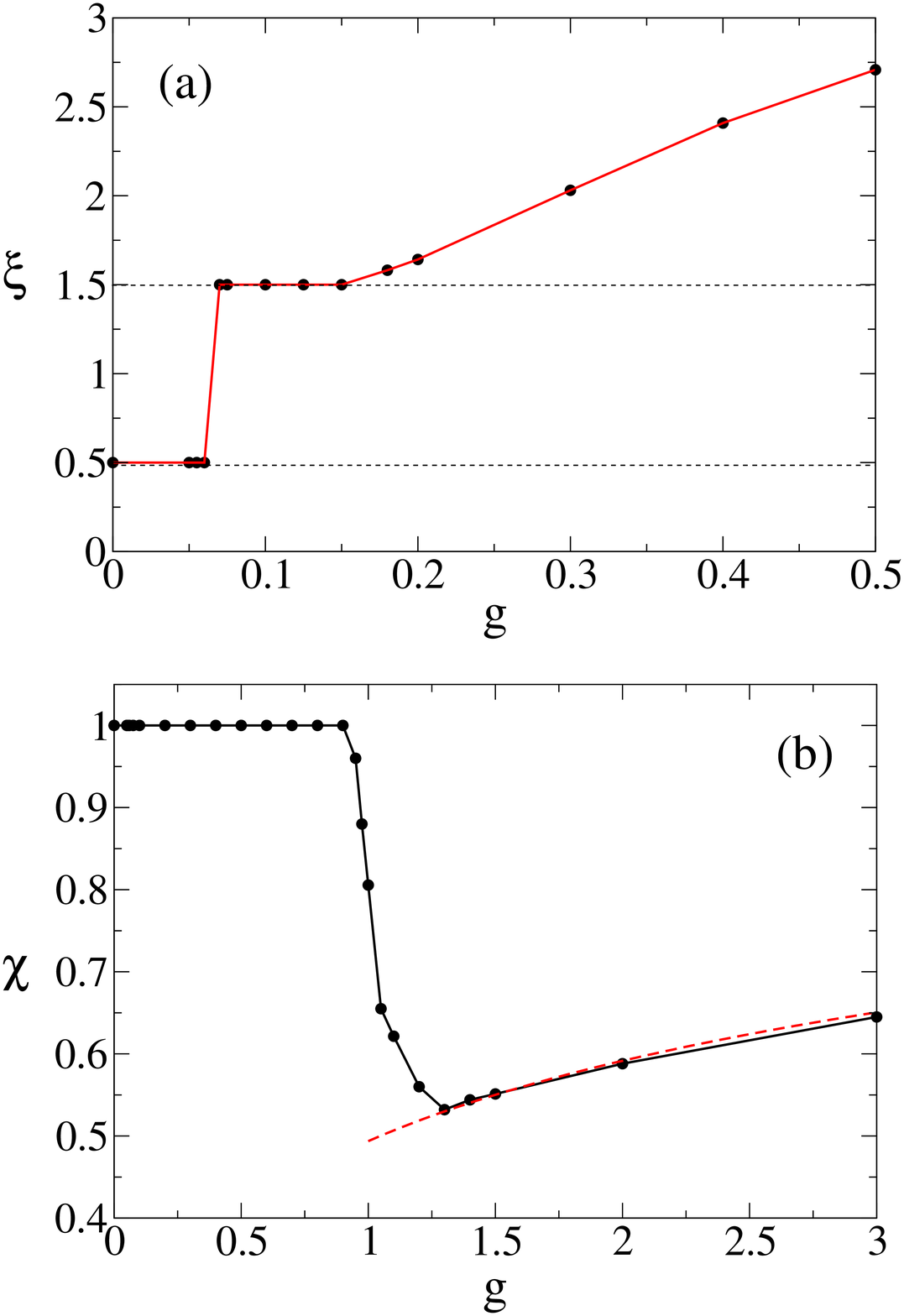}
\vspace*{-0.2cm}
\caption{(a) Value of $\xi=\langle |l+1/2|\rangle $ as a function of $g$ for $\kappa=15$, 
the dashed lines indicate the value $1/2$ and $3/2$ characteristic of HV(1/2) and HV(3/2), 
respectively; (b) Aspect ratio $\chi$ of the cloud for the same case. The 
dashed curve indicates the expected aspect ratio from Eq.~\eqref{eq:Var}.}
\label{fig:2}
\end{center}
\vspace*{-0.7cm}
\end{figure}

Decomposing $\psi({\bf k})=\sum a_l \Psi_l({\bf k})$, we evaluate 
$\xi\equiv\langle |l+1/2|\rangle =\sum |a_l|^2 |l+1/2|$~(Fig.~\ref{fig:2}(a)).  
For $g<g_{cr}$, the phase HV(1/2) is characterized by a plateau at $\xi=1/2$. 
At $g=g_{cr}$ a sudden jump occurs into $\xi=3/2$. Interestingly, the numerical 
simulation of Eq.~(\ref{eq:GPE}) shows that up to a second critical $g_{cr}^{(2)}$, 
a second plateau at $\xi=3/2$ occurs. We denote this phase as HV(3/2), 
which is characterized by a ring-like density profile~(Fig.~\ref{fig:1}(b)). As for the HV(1/2), $l=1$ and $l=-2$ are degenerated, 
and any linear combination of them produces the same rotationally symmetric total density profile. 
Again, as for HV(1/2), the selection of $l=1$ or $l=-2$ leads to a ring-like momentum distribution
whereas combinations of them lead to a $\cos(3\phi)$ modulation along the ring.
%%%%%%%%%%%%%%%%%%
%% FIGURE 2

%%%%%%%%%%%%%%%%%%%%%%%%%%%%%%%%%%%%%%%%

\paragraph{Lattice phase.}
For $g>g_{cr}^{(2)}$, $\xi$ departs from the $3/2$ plateau, rotational symmetry is broken, 
and a different hexagonally-symmetric phase occurs, characterized by the developing of 
a triangular lattice of minima in the total density profile~(Fig.~\ref{fig:1}(c)), which resembles 
that observed for vortex lattices in rapidly-rotating condensates~\cite{Raman2001}.
The spacing between the minima is independent of $g$, being $\simeq 4\pi/3\kappa$, such that the 
border of the first Brillouin zone of the lattice lies on the $k=\kappa$ ring.
On the contrary, as discussed below, the width of the cloud envelope $\sqrt{\langle x^2\rangle}$ is independent of $\kappa$ for dominant SOC, 
being only dependent on $g$. Hence enhancing interactions just increases the number of observed density minima ($\propto\kappa^2\langle x^2\rangle$) keeping 
invariant the lattice structure. The individual components present in typical numerical simulations an involved density and phase 
distribution, characterized by vortices and anti-vortices of different quantizations. In this Letter 
we are not interested in them, since the particular distribution among the components may 
depend on details of spin-dependent interactions.

An interesting insight about the lattice phase may be gained from 
its momentum distribution characterized by the appearance of six maxima 
at angles $\varphi_j=j\pi/3$, around the ring of radius $\kappa$~(Fig.~\ref{fig:1}(d)).
The lattice hence results from the combination of three pairs of opposite momenta.
Major features of the overall density profile may be obtained from these six-peaked structure by means of a 
Gaussian ansatz of the form:
\begin{equation}
\psi({\bf k})\!=\!\sum_j \!a_j 
e^{-\frac{1}{2}\left[l_r^2(k_r^{(j)}-\kappa)^2+l_\varphi^2 k_\varphi^{(j)^2}\right ]}
\!\!\left(\!\! 
\begin{array}{c} 
e^{-\frac{i}{2}(\varphi_j+k_\varphi^{(j)}/\kappa)} \\ 
e^{\frac{i}{2}(\varphi_j+k_\varphi^{(j)}/\kappa)}
\end{array} 
\!\!\right)
\label{eq:ansatz}
\end{equation}
with $a_j=1/\sqrt{6}$, $k_r^{(j)}=k_x\cos\varphi_j + k_y\sin\varphi_j$ and 
$k_\varphi^{(j)}=-k_x\sin\varphi_j + k_y\cos\varphi_j$. 
This solution is characterized 
by $\langle x^2\rangle=\langle y^2\rangle=(l_r^2+l_\varphi^2)/4$. 
For a sufficiently large $\kappa$, the energy may be approximated as 
$E(\eta,g)=2\langle x^2\rangle=\frac{1}{2}\sqrt{(1+\eta^2)(1+g f(\eta)/3\pi)}$, 
with $f(\eta)=1/\eta+8/\sqrt{3+10\eta^2+3\eta^4}$, and $\eta=l_\varphi/l_r$. Minimizing the energy 
we obtain $\langle x^2\rangle$ in excellent agreement with our numerical simulations 
(since the orientation of the cloud is arbitrary, the $x$ direction is determined numerically as that with the larger width).
Interestingly, the kinetic energy $E_{kin}=1/4l_r^2$ just contains the radial width $l_r$, 
since the $l_\varphi$ dependence is exactly cancelled by the SOC. This cancellation leads to 
a global shrinking of the cloud for the lattice phase, and plays 
a major role in the stripe phase discussed below.

\paragraph{Stripe phase.}
All the phases mentioned above fulfill $\chi^2\equiv \langle y^2 \rangle / \langle x^2 \rangle =1$, either due to a full rotational 
symmetry or due to hexagonal symmetry in the case of the lattice. However, at a sufficiently large $g$, 
the ratio $\chi$ departs from $1$~(Fig.~\ref{fig:2}(b)), i.e. the cloud breaks the hexagonal symmetry, becoming elongated along $x$~(Fig.~\ref{fig:1}(e)). 
After a transient, in which $\chi$ decreases abruptly~(Fig.~\ref{fig:2}(b)), the condensate acquires for a sufficiently large 
$g$ a stripe form along the major axis $x$~(Fig.~\ref{fig:1}(f)), similar to that obtained in the case of homogeneous condensates with SOC~\cite{Wang2010}. 

The properties of the trapped stripe, and in particular its elongation along the direction 
of the stripe modulation, may be understood from the ansatz~\eqref{eq:ansatz}, but just considering 
two peaks on the ring at $\varphi_j=0,\pi$. In this case, $\langle x^2 \rangle =l_r^2/2$ and 
$\langle  y^2 \rangle=l_\varphi^2/2$. For a sufficiently large $\kappa$ 
the energy functional may be approximated as
\begin{equation}
E\simeq\frac{1}{4l_r^2}+\frac{l_r^2+l_\varphi^2}{4}+\frac{g}{4\pi l_r l_\varphi}. 
\end{equation}
As mentioned above, the kinetic energy, which in absence of SOC is of the form $\frac{1}{4lr^2}+\frac{1}{4l_\varphi^2}$,
reduces to $\frac{1}{4l_r^2}$, since the $l_\varphi$ contribution is exactly cancelled by the SOC. 
This breaks the symmetry between $x$ and $y$ axis, leading to an effective compression along $y$.  
Energy minimization leads to an equation for the aspect ratio 
\begin{equation}
\frac{g}{2\pi}=\frac{\chi^3}{1-\chi^2},
\label{eq:Var}
\end{equation}
which provides an excellent agreement with our numerical calculations as shown in Fig.~\ref{fig:2}(bottom).

\begin{figure}%[ht]
%\vspace{-0.7cm}
\begin{center}
\includegraphics[width=0.45\textwidth,angle=0]{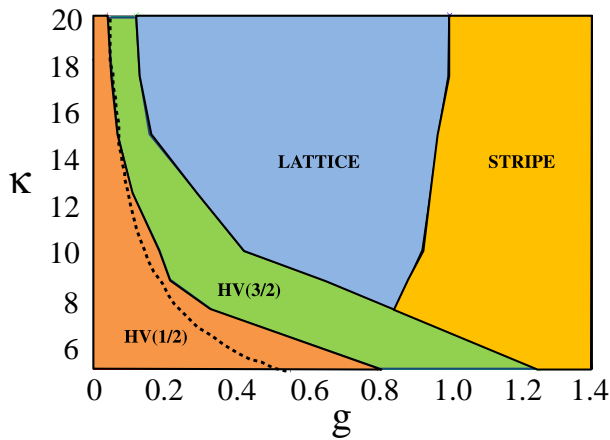}
\vspace{-0.3cm}
\caption{Phase diagram as a function of $\kappa$ and $g$. The dashed line indicates the result of
Eq.~\eqref{eq:gcr}.}
\label{fig:3}
\end{center}
\vspace*{-0.7cm}
\end{figure}

\paragraph{Phase diagram.}
We have obtained the ground-state phase diagram by numerically solving Eq.~\eqref{eq:GPE} for a large 
number of $g$ and $\kappa$ values. Our results are summarized in Fig.~\ref{fig:3}. 
The border $g_{cr}^{(3)}$ between the lattice and the stripe phase is determined as the $g$ value at which 
$\chi\simeq 0.75$. Due to  the $1/\kappa^2$ dependence of the angular modes, for large $\kappa$ 
values the HV(1/2) and HV(3/2) shrink, being confined to a regime of 
progressively smaller $g$ values as $\kappa$ increases. On the contrary,  
the lattice phase becomes progressively more favorable extending for very large $\kappa$ almost to $g=0$. 
The border between lattice and stripe approaches $g\simeq 1$ for large $\kappa$.
When $\kappa$ decreases, the lattice phase shrinks until disappearing below 
$\kappa_c\simeq 7.5$. Below $\kappa_c$ there is a direct HV(3/2) to stripe transition, at which 
the rotational symmetry is broken. 

%%%%%%%%%%%%%%%%%%
%% FIGURE 3

%%%%%%%%%%%%%%%%%%%%%%%%%%%%%%%%%%%%%%%%

\paragraph{Conclusions.}
In summary trapped spin-orbit coupled BECs present intriguing ground-state properties as a result 
of the interplay between trap energy, spin-orbit coupling and interactions. The corresponding 
phase diagram includes two rotationally symmetric phases, HV(1/2) and HV(3/2), a hexagonally-symmetric phase with a triangular 
lattice of minima, which resembles a vortex lattice, and a stripe phase, similar to that 
predicted for homogeneous BECs. In experiments the phases should be clearly distinguisable either 
by monitoring in situ the density profile, or by time-of-flight imaging. The latter maps the momentum distribution, 
and, in particular, the lattice phase will be characterized by six out-going clouds.

Finally, let us point that in our paper we have considered for simplicity spin-independent interactions. 
In typical experiments spin-dependent interactions are much weaker than spin-independent ones, 
and hence should not affect the overall density 
profiles, although the particular distribution among the components will change. However, 
due to the dressed nature of the states in actual experiments, spin-dependent interactions do not 
simply reduce to a different inter-component density-density interaction, as in usual two-component 
condensates. On the contrary, phase-dependent terms will occur~\cite{Zhang2011}, which may considerably complicate and enrich 
the physical picture. This issue, and anisotropic effects, both in the trap and in the SOC, will be the focus of further investigation.

{\em Note added:} When completing this paper we became aware of two recent works in which a 
similar problem is considered. In Ref.~\cite{Hu2011}, which considers the effect of spin-dependent interactions, 
phases IIA (and IA), IA' (and part of IIB) and IB (and part of IIB) correspond to the HV(1/2), HV(3/2) and lattice phases. 
A new recent version of Ref.~\cite{Wu2008} discusses the border of the stripe phase at small $\kappa<4$, 
showing a direct half-vortex to stripe transition, in agreement with our results. 

% Acknowledgements
We acknowledge support from the Center for Quantum Engineering and Space-Time Research QUEST.

% -------------------------------------------------------------------------------------------

\end{document}